\documentclass[11pt,twoside]{article}
\usepackage{asp2010}
\resetcounters

\begin{document}

\title{The Unified Astronomy Thesaurus}
\author{Alberto Accomazzi$^1$, Norman Gray$^2$, 
  Chris Erdmann$^1$, Chris Biemesderfer $^3$, Katie Frey $^1$, and Justin Soles $^4$
\affil{$^1$Harvard-Smithsonian Center for Astrophysics, 60 Garden Street, Cambridge, MA 02138, USA}
\affil{$^2$School of Physics and Astronomy, University of Glasgow, Glasgow, G12 8QQ, UK}
\affil{$^3$American Astronomical Society, 2000 Florida Avenue, NW, Washington, DC 20009, USA}
\affil{$^4$School of Information Studies, McGill University, 3661 Peel Street, Montreal, Qc H3A 1X1}}

\begin{abstract}
The Unified Astronomy Thesaurus (UAT) is an open, interoperable and community-supported 
thesaurus which unifies the existing divergent and isolated Astronomy \& Astrophysics vocabularies
into a single high-quality, freely-available open thesaurus formalizing astronomical 
concepts and their inter-relationships.  The UAT builds upon the existing IAU Thesaurus 
with major contributions from the astronomy portions of the thesauri developed by the 
Institute of Physics Publishing, the American Institute of Physics, and SPIE.
We describe the effort behind the creation of the UAT and the process through
which we plan to maintain the document updated through broad community participation.
\end{abstract}

\section{History}

In astronomy, there have been different initiatives aimed at creating classification
systems to be used in the literature.  In 1992, Shobbrook and Shobbrook published
a 2,551-term thesaurus which was endorsed by the IAU \citep{1992PASAu..10..134S} and named
the IAU Thesaurus.  
Meanwhile, editors from the main astronomy journals developed a parallel system of keywords
to characterize and indexed published articles called ``Astronomy Subject Headings'' 
(or more commonly simply ``journal keywords'')
which was adopted by the journals also in 1992.  This system, consisting of just over 
300 concepts organized in a hierarchy, was considered simpler to use than the IAU 
Thesaurus and appropriate for creating the annual subject heading index used for
browsing the content of the journals.  With the passing of time, limitations of this
system have become apparent, yet it continues to be used today in all the major astronomy
publications.
In 2007 the International Virtual Observatory Alliance (IVOA) Semantics Working Group
published a study on 
the use of Vocabularies in the Virtual Observatory (\cite{2009ASPC..411..179G} 
and \cite{2011arXiv1110.0520D}), providing versions of the IAU Thesaurus and the
astronomy subject headings in SKOS (Simple Knowledge Organization System, \citet{skosref}) format.

Outside of astronomy, the American Institute of Physics (AIP), on behalf of the publishers
in the field, developed a more comprehensive classification system which included
astronomy terms.  Named PACS (Physics and Astronomy Classification System), the 
keywords were originally proposed in 1975 and have been used 
to characterize content in most of the major
physics journals until 2011, when AIP
announced that they would stop maintaining
and using this classification scheme in favor of a more modern system.
With the end of PACS, the physics community found itself without a 
common system for classifying new journal articles, and in the
fall of 2011 a group of physics and astronomy publishers met 
to explore the possibility of joining forces in developing a
modern system to support the classification and semantic enrichment of the literature
(the new thesaurus's lineage is illustrated in Figure 1).
Participants 
included members of the astronomical community involved in the publications and curation
journal articles, including representatives from ADS, the CfA Wolbach 
Library, and the Institute of Physics (IOP), the publisher of AAS journals.

Recognizing the danger of having different publishers develop separate thesauri
based on the content that they managed, a key group of participants at the meeting
felt that there was an opportunity to collaborate on the development of a single system
covering at least astronomy and astrophysics.
IOP and AIP, in collaboration with ADS, the CfA Wolbach Library and the IVOA Semantics working group
began discussing the possibility of working together on a single astronomy thesaurus
created by merging and reconciling existing and emerging vocabularies and thesauri.
As the plans to create a unified thesaurus were taking shape, the American Astronomical
Society (AAS) joined the effort and provided logistical and legal support helping 
negotiate the licensing terms for the final work.  
Having settled issues related to intellectual rights in the fall of 2012,
AIP and IOP proceeded to donate 
the astronomy portions of their thesauri and funded an effort to merge them with the
IVOA-enhanced version of the IAU thesaurus.  The resulting work, newly named the Unified
Astronomy Thesaurus, was born.

\begin{figure}[!ht]
\plotone{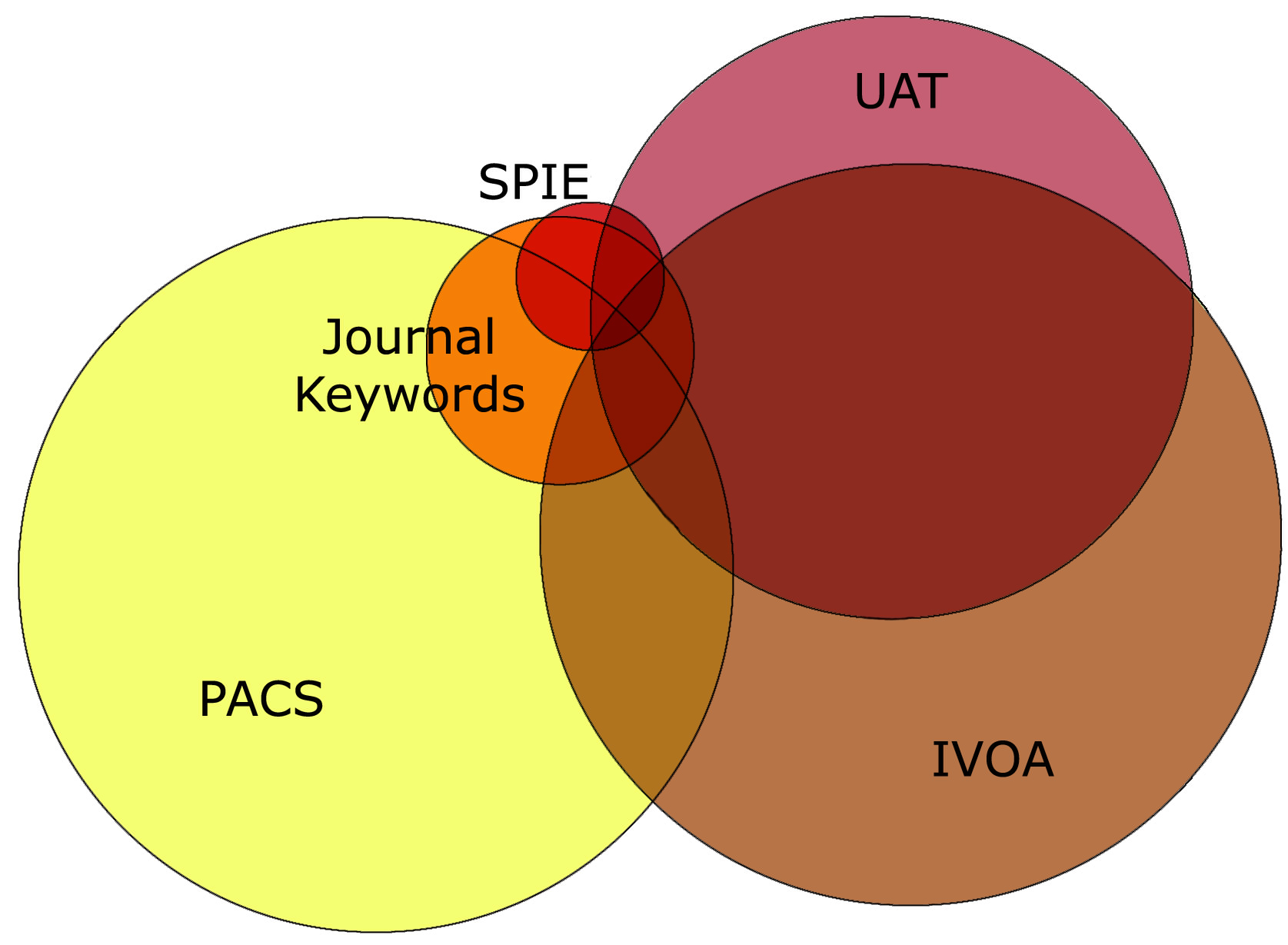}
\caption{Lineage of the Unified Astronomy Thesaurus: the diagram illustrates the relative size 
and overlap of the concepts which contributed to the list of terms in the UAT.}
\end{figure}

\section{Current Status}

The Unified Astronomy Thesaurus is available under
a Creative Commons License (CC-BY-SA), ensuring its widest use while protecting the intellectual 
property of the contributors.  While it is being managed by the original core group of
people, we envision that its development and 
maintenance will be stewarded by a broader group of parties having a direct stake in it.  
This includes professional associations (IVOA, IAU), learned societies (AAS, RAS), 
publishers (IOP, AIP), librarians and other curators working for major astronomy 
institutes and data archives.

While the main impetus behind the creation of a single thesaurus has been the wish to 
support semantic enrichment of the literature, we expect that use of the UAT 
(along with other vocabularies and ontologies currently being developed in our community) 
will be much broader and will stimulate the development of a wide range of 
astronomy resources, including data products and services.

There are a number of differences which distinguish the UAT from
the prior efforts in this area, and which are worth mentioning:
{\bf Openness}: the UAT is licensed under liberal terms, and can therefore be easily
integrated in workflows and applications at no cost.  Additionally, anyone interested in astronomy can 
contribute to the UAT by suggesting additions, refinements, revisions and modifications to it.
{\bf Interoperability}: the UAT is available as a SKOS
document, a standard open format which facilitates its embedding and re-use in larger contexts.
{\bf Community support}: The UAT is supported by major stake-holders in the astronomical community,
including journal publishers, professional societies, libraries and archives, and VO-related projects.

\section{Curation and Development}

Astronomy and Astrophysics are relatively fast-changing disciplines,
where new theories and discoveries (like dark energy and exoplanets) 
spawn new research fields on a regular basis.  Since one of the main
goals of the UAT is to provide a formal language that can be used to 
describe this entire field, we need to ensure that the UAT remains a
living document, regularly updated and revised to capture any new
concepts being introduced in the scholarly literature.
Here is where we hope technology will come to our aid:
contrary to some of the prior efforts which required a great amount of
manual work to review all the published literature, 
we plan to use text mining tools
to detect and suggest new terms for inclusion as they appear in new papers.  
The process supporting this activity is currently being developed by publishers
and ADS to discover topics discussed in papers and classify them appropriately.

In our effort to make the community aware of the efforts behind the UAT,
in late 2012 we launched a website to provide information on the status of the thesaurus 
and as a way to disseminate it.  The home of the UAT is now: \url{http://astrothesaurus.org}.  
In addition to the most recent release of the thesaurus, the website provides a
browsable interface allowing users to view the concepts in the thesaurus and their relationships.
We encourage all interested parties to stay up to date by accessing the information available on the 
website and by participating in the on-line discussion on the UAT mailing list
(subscription information also available from the website).
The thesaurus is currently being reviewed for accuracy and consistency while we finalize
the curatorial process, with a full release scheduled in 2014.

On the grounds that the best curation of a new thesaurus can only be provided by the community itself, 
we intend to enhance the existing web interface to allow not only the UAT to be browsed, but also 
enable anyone to identify and suggest additions, changes and improvements to terms within the UAT. 
These suggestions will be filtered through volunteer editors (astronomers or subject matter experts) 
and entered into the UAT with credit to the contributor and editor. Periodically, a librarian will 
review and tune the overall structure before re-releasing it, allowing the editors to focus 
solely on reviewing the suggestions made by users. The involvement of the astronomical community 
-- with the development work led by the CfA Wolbach Library and ADS -- will help the UAT remain an up-to-date, 
accurate and trusted resource for the community as a whole. 

\acknowledgements Support for the operations of ADS is provided by NASA under Grant NNX12AG54G

\bibliography{P60}

\end{document}